\documentclass{article}
\usepackage{amsfonts}

\usepackage{amscd,amsmath,amssymb}

\newcommand{\ba}{\begin{array}}
\newcommand{\ea}{\end{array}}
\newcommand{\be}{\begin{equation}}
\newcommand{\ee}{\end{equation}}
\newcommand{\bea}{\begin{eqnarray}}
\newcommand{\eea}{\end{eqnarray}}
\newcommand{\bi}{\begin{itemize}}
\newcommand{\ei}{\end{itemize}}

\begin{document}

\thispagestyle{empty}

\begin{center}
{~}\\[0pt]
\vspace{3cm} {\Large \textbf{``Root'' Action for }}{\Large $\mathcal{N}$}
{\Large \textbf{=4 Supersymmetric Mechanics Theories }}\vspace{2cm}\\[0pt]
{\large \textbf{S.~Bellucci${}^{a}$, S.~Krivonos${}^{b}$,
A.~Marrani$^{c,a}$
and E.~Orazi${}^{a,d}$}}\\[0pt]
\vspace{2cm} \textit{${}^{a}$INFN-Laboratori Nazionali di Frascati,
C.P. 13,
00044 Frascati, Italy}\\[0pt]\texttt{bellucci, marrani, orazi@lnf.infn.it} \\[0pt]
\vspace{0.5cm} \textit{${}^{b}$ Bogoliubov Laboratory of Theoretical
Physics, JINR, 141980 Dubna, Russia}\\[0pt]
\texttt{krivonos@theor.jinr.ru} \\[0pt]
\vspace{0.5cm} \textit{${}^{c}$Centro Studi e Ricerche ``Enrico
Fermi'', Via Panisperna 89A, 00184 Roma, Italy }\\[0pt]
\vspace{0.5cm} \textit{${}^{d}$ Dipartimento di Fisica,
Universit\`{a} di
Roma Tor Vergata, Via della Ricerca Scientifica 1, 00133 Roma, Italy}\vspace{%
2.5cm}

\textbf{Abstract}
\end{center}

We propose to consider the $\mathcal{N}=4,d=1$ supermultiplet with $%
(\mathbf{4,4,0})$ component content as a ``root'' one. We elaborate a new
reduction scheme from the ``root'' multiplet to supermultiplets with a
smaller number of physical bosons. Starting from the most general
sigma-model type action for the ``root'' multiplet, we explicitly
demonstrate that the actions for the rest of linear and nonlinear $%
\mathcal{N}=4$ supermultiplets can be easily obtained by reduction.

Within the proposed reduction scheme there is a natural
possibility to introduce Fayet-Iliopoulos terms. In the reduced
systems, such terms give rise to potential terms, and in some
cases also to terms describing the interaction with a magnetic
field.

We demonstrate that known $\mathcal{N}=4$ superconformal actions, together
with their possible interactions, appear as results of the reduction from a
free action for the ``root'' supermultiplet. As a byproduct, we also
construct an $\mathcal{N}=4$ supersymmetric action for the linear $(\mathbf{%
3,4,1})$ supermultiplet, containing both an interaction with a
Dirac monopole and a harmonic oscillator-type potential,
generalized for arbitrary conformally flat metrics. \hfil
\newpage

\section{Introduction}

One of the most interesting features of the $\mathcal{N}=4$
supersymmetric mechanics theories, which makes them rather
different from their
higher dimension counterparts, is the existence of irreducible $\mathcal{N%
}=4,d=1$ supermultiplets with all possible numbers of physical bosons:
starting from a supermultiplet with four physical bosons $(\mathbf{4,4,0})$
and finishing with the supermultiplet $(\mathbf{0,4,4})$, containing no
physical bosons at all\footnote{%
We are using the notation $(\mathbf{m,4,4-m})$ identifying an off-shell $%
\mathcal{N}=4,d=1$ supermultiplet with $\mathbf{m}$ physical bosons, $%
\mathbf{4}$ fermions and $\mathbf{4-m}$ auxiliary bosonic components.}. Some
of these supermultiplets can be obtained by dimensional reduction from,
say, four dimensional superfield theories. Other supermultiplets exist only
in one dimension.

It was shown by J.~Gates, Jr. and L.~Rana \cite{GR} that (at least linear) $%
\mathcal{N}=4,d=1$ supermultiplets are closely related among themselves. It
turns out that in one dimension one may switch between different
supermultiplets by a proper expression of auxiliary bosonic components
through time-derivatives of physical bosons, and vice versa. In \cite{ikl1}
a wide set of off-shell $\mathcal{N}=4$ supermultiplets with \textbf{4}
physical fermions and a finite number of auxiliary fields was deduced by
using nonlinear realizations of the unique $\mathcal{N}=4,d=1$
superconformal group $D(2,1;\alpha )$. Besides already known supermultiplets
\cite{{GR},{FR},{leva},{is},{BP}}, two new off-shell nonlinear multiplets
were found. This approach shed some light on the geometric nature of the
relations between different supermultiplets. Indeed, it was shown that the
physical bosonic components of the obtained $\mathcal{N}=4,d=1$
supermultiplets parameterized different coset spaces of the same
superconformal group. Therefore, the relations between supermultiplets arise
as a result of switching between the corresponding cosets.

Being completely off-shell, the relations between $\mathcal{N}=4$
supermultiplets have nothing to do with the actions describing the
various versions of supersymmetric mechanics. Thus, a natural
question arises: \textit{how general are such relations?}
Otherwise speaking, one is naturally led to ask whether it is
possible to start from the most general sigma-model type action
for one supermultiplet and then, by using the proper relations,
get the general action for another multiplet. A related question
concerns the restrictions that should be imposed on the metrics of
the bosonic manifolds, in order to succeed with such a reduction.
In this respect, the supermultiplet $(\mathbf{4,4,0})$ with the
maximum number of physical bosons should be distinguished as a
``root'' supermultiplet\footnote{The term ``root supermultiplet'' has been proposed
in \cite{FG}.}, because the corresponding metric of the
bosonic manifold depends on all four bosons. When passing to
supermultiplets with a smaller number of physical bosons, clearly
such a metric has to be properly restricted to depend only on the
physical bosons ``surviving'' after the reduction. One should also
point out that the considerations in \cite{GR} were purely
algebraic, without any reference to the automorphisms. But if some
specific action is considered when using the relations between
different supermultiplets, then one should pay attention to the
corresponding automorphism symmetries.

In the present paper we answer some of the above mentioned questions.
Starting from the most general sigma-model type action for the ``root'' $(%
\mathbf{4,4,0})$ multiplet, we explicitly demonstrate that the actions for
the rest of linear, and two recently-discovered nonlinear, known $\mathcal{N}%
=4$ supermultiplets can be easily obtained by a proper reduction
procedure. We propose a particular parametrization of the
``root'' supermultiplet, with a clear geometric meaning
\cite{ikl1}, and suitable for reduction procedure. Within the
proposed reduction scheme there is the possibility to introduce
Fayet-Iliopoulos (FI) terms. Such terms, when properly treated in
the reduction procedure, generate the interactions.

One of the most interesting results we achieve is that known
$\mathcal{N}=4$ superconformal actions, together with their
possible interactions, appear as results of the reduction from a
\textit{free action} for the ``root'' supermultiplet. As a
byproduct, we also construct an $\mathcal{N}=4$ supersymmetric
mechanics action for the linear $(\mathbf{3,4,1})$ supermultiplet,
interestingly containing both an interaction with a Dirac monopole
and a harmonic oscillator-type potential, generalized for
arbitrary conformally flat metrics.

Summarizing, $\mathcal{N}=4,d=1$ supersymmetry is found to be characterized
by the existence of a ``root'' supermultiplet with a related ``root'' action
from which, by performing a proper reduction procedure, one may construct
the rest of the supermultiplets with the corresponding actions.

\setcounter{equation}0

\section{The ``root'' supermultiplet and its action}

In order to describe the $\mathcal{N}=4$ supermultiplet with four bosonic
and four fermionic physical components, we follow \cite{{ikl1},{il}}, by
introducing a real quartet of $\mathcal{N}=4$ superfields $\mathcal{Q}%
^{ia}\;\left( (\mathcal{Q}^{ia})^{\dagger }=\mathcal{Q}_{ia}\right) $,
depending on the coordinates of the $\mathcal{N}=4,d=1$ superspace $\Bbb{R}%
^{(1|4)}$
\begin{equation}
\Bbb{R}^{(1|4)}=(t,\theta ^{iA})\,,\qquad \left( \theta
^{iA}\right) ^{\dagger }=\theta _{iA}\,,\nonumber
\end{equation}
where $(i,a,A=1,2)$ are doublet indices of three commuting $SU(2)$
subgroups. The constraints which identify the multiplet are\footnote{%
We use the following convention for the skew-symmetric tensor $\epsilon $: $%
\epsilon _{ij}\epsilon ^{jk}=\delta _{i}^{k}$, $\epsilon
_{12}=\epsilon ^{21}\equiv1$. As usually, the round brackets
express symmetrization of the enclosed indices.}
\begin{equation}
D_{A}^{(i}\mathcal{Q}^{j)a}=0,  \label{root1}
\end{equation}
where the covariant spinor derivatives $D^{iA}$ are defined by
\begin{equation}
D^{iA}\equiv \frac{\partial }{\partial \theta _{iA}}+i\theta ^{iA}\partial
_{t}\,,\quad \quad \left\{ D^{iA},D^{jB}\right\} =2i\epsilon ^{ij}\epsilon
^{AB}\partial _{t}~.  \label{sder}
\end{equation}
Since the defining constraints (\ref{root1}) imply that
\begin{equation}
D^{iA}D^{jB}\mathcal{Q}{}^{ka}=2i\epsilon ^{AB}\epsilon ^{kj}\partial _{t}%
\mathcal{Q}{}^{ia},
\end{equation}
one may immediately conclude that the independent components of the
superfields $\mathcal{Q}^{ia}$ are only\footnote{%
From the reality of superfields $\mathcal{Q}{}^{ia}$ it follows
that $\left( \psi ^{aA}\right) ^{\dagger }=-\psi _{aA}$. }
\begin{equation}
q^{ia}\equiv \mathcal{Q}^{ia}|_{\theta _{iA}=0},\;\quad \quad \psi
^{aA}\equiv \frac{1}{2}D_{i}^{A}\mathcal{Q}^{ia}|_{\theta _{iA}=0},
\label{comp}
\end{equation}
and thus $\mathcal{Q}^{ia}$ describe the irreducible (\textbf{4,4,0})
supermultiplet. This supermultiplet was considered in the component and $%
\mathcal{N}=1$ superfield approaches in \cite{mult}; it was also recently
studied in \cite{{ikl1},{il}} by using the $\mathcal{N}=4$ superspace
formalism.

The most general $\mathcal{N}=4$ supersymmetric $\sigma $-model type action
for the superfields $\mathcal{Q}{}^{ia}$ constrained by conditions (\ref
{root1}) may be written in the full $\mathcal{N}=4,d=1$ superspace as
\begin{equation}
S=\int dtd^{4}\theta L(\mathcal{Q})\;,  \label{root2}
\end{equation}
where $L(\mathcal{Q})$ is an arbitrary scalar function. Passing to the
components in the action (\ref{root2}) we get
\begin{equation}
S=\int dt\left[ G\left( \dot{q}^{ia}\dot{q}_{ia}+\frac{i}{2}\psi ^{aA}\dot{%
\psi}_{aA}\right) +\frac{i}{2}\frac{\partial G}{\partial
q^{ia}}\psi ^{aA}\psi
_{A}^{b}\;\dot{q}_{b}^{i}-\frac{1}{48}\bigtriangleup G\,\psi
^{aA}\psi _{A}^{b}\psi _{a}^{B}\psi _{Bb}\right] ,  \label{root3}
\end{equation}
where the metric $G$ is defined as
\begin{equation}
G\equiv \bigtriangleup L=\frac{\partial ^{2}}{\partial q^{ia}\partial q_{ia}}%
L(q)\;.
\end{equation}
The component action (\ref{root3}) is invariant under off-shell $\mathcal{N}%
=4$ supersymmetry realized as ($\left( \varepsilon ^{iA}\right) ^{\dagger
}=\varepsilon _{iA}$)
\begin{equation}
\delta q^{ia}=-\varepsilon ^{iA}\psi _{A}^{a}~,\quad \delta \psi
^{aA}=2i\varepsilon ^{iA}\dot{q}_{i}^{a}~.  \label{root4}
\end{equation}

Due to the absence of bosonic auxiliary components, potential terms cannot
be added to the action (\ref{root2}). Nevertheless, one may still modify the
action (\ref{root2}) by a adding a self-interacting term
\begin{equation}
S_{int}\equiv\frac{1}{72}\int dtd^{4}\theta \;A_{(ab)}\theta ^{iA}\theta _{A}^{j}%
\mathcal{Q}{}_{i}^{a}\mathcal{Q}{}_{j}^{b}.  \label{int}
\end{equation}
Here, the real constants $A_{(ab)}$ are symmetric in the indices and
their appearance in the action (\ref{int}) explicitly breaks the
corresponding $SU(2)$ invariance down to $U(1)$. In the components
(\ref {comp}) the action (\ref{int}) reads
\begin{equation}
S_{int}\equiv\int dt\left( 2i{\dot{q}}{}_{i}^{a}q^{ib}+\psi
^{Aa}\psi _{A}^{b}\right) A_{(ab)}.  \label{int1}
\end{equation}
As it may easily be seen, the action (\ref{int})-(\ref{int1}) describes the
interaction of our supermultiplet with the magnetic field. More general
self-interacting terms may be also constructed \cite{il}, but we will not
consider them in what follows.

\setcounter{equation}0

\section{From the ``root'' supermultiplet to the (3,4,1) ones}

Since the appearance of \cite{GR}, it is well known that the linear $%
\mathcal{N}$-extended supermultiplets in $d=1$ are deeply related among
themselves. Roughly speaking, the transformation properties of any bosonic
auxiliary component $\mathcal{A}$ of the $(\mathbf{m,}\mathcal{N}\mathbf{,}%
\mathbf{\mathcal{N}-m})$ supermultiplet read
\begin{equation}
\delta \mathcal{A}\sim \mbox{parameter}\times \partial _{t}\left(
\mbox{
physical fermions}\right) .  \label{disc1}
\end{equation}
In $d=1$ we may easily integrate (\ref{disc1}) and pass to the physical
bosonic components $\Phi $ defined as
\begin{equation}
\partial _{t}\Phi =\mathcal{A}  \label{disc2}
\end{equation}
with the supersymmetry transformation properties
\begin{equation}
\delta \Phi \sim \mbox{parameter}\times \left( \mbox{ physical
fermions}\right) .  \label{disc3}
\end{equation}
Therefore, such an integration procedure amounts to replace the auxiliary
bosonic field $\mathcal{A}$ with the physical bosonic field $\Phi $, and
consequently the resulting new supermultiplet is the $(\mathbf{m+1,}\mathcal{%
N}\mathbf{,}\mathcal{N}\mathbf{-m-1})$ one. It is clear that one may reverse
the consideration and pass from (\ref{disc3}) to (\ref{disc1}) or, in other
words, from the $(\mathbf{m,}\mathcal{N}\mathbf{,}\mathcal{N}\mathbf{-m})$
supermultiplet to the $(\mathbf{m-1,}\mathcal{N}\mathbf{,}\mathcal{N}\mathbf{%
-m+1})$ one. In this way one may recover all $\mathcal{N}$-extended
supermultiplets starting from the basic, ``root'' supermultiplet, which
contains no auxiliary components at all. We have to choose such a multiplet
as the ``root'' one, because the metric of the corresponding bosonic
manifold will depend on the maximal number of physical bosons.

Although being intuitively transparent, the discussed procedure leaves some
questions unanswered.

First of all, the straightforward change of some physical bosons
into auxiliary ones will destroy most of the automorphisms of the
original system. Thus, it would be nice to exploit an approach
preserving as many symmetries of the ``root'' system as possible.

However, the most important question concerns the possibility to
generate potential terms by the ``flipping'' of some physical
bosonic components into auxiliary ones. Interestingly, we are
going to demonstrate that this can actually be done, and that most
of the physically interesting potential terms simply come from the
FI terms introduced during the reduction.

It should be also mentioned that the above considered relations between
different supermultiplets are purely kinematical, and therefore it would be
interesting to see how all these transitions between supermultiplets work
dynamically, i.e. at the level of the corresponding action.\smallskip

In this Section we will reduce the $\mathbf{(4,4,0)}$ $\mathcal{N}=4,d=1$
``root'' supermultiplet $\mathcal{Q}{}^{ia}$ to some supermultiplets with a
smaller number of physical bosons, namely to the $\mathbf{(3,4,1)}$ ones. In
order to preserve all possible automorphisms during the reduction procedure,
it is useful to define new bosonic variables as follows \cite{ikl1}:
\begin{equation}
q^{11}\equiv \frac{e^{\frac{1}{2}\left( u-i\phi \right) }}{\sqrt{1+\Lambda
\overline{\Lambda }}}\Lambda ~,\quad q^{21}\equiv -\frac{e^{\frac{1}{2}%
\left( u-i\phi \right) }}{\sqrt{1+\Lambda \overline{\Lambda }}}~,\quad
q^{22}\equiv \overline{\left( q^{11}\right) }~,\quad q^{12}\equiv -\overline{%
\left( q^{21}\right) }~.  \label{newvar}
\end{equation}
It is worth pointing out that the definitions (\ref{newvar}) have
a clear geometric meaning. Indeed, $e^{u}$ plays the role of
radius, while the variables $\Lambda ,\bar{\Lambda}$ parameterize
the 2-sphere $SU(2)/U(1)$, with the angular variable $\phi $ being
responsible for $U(1)$ rotations. Thus, the most symmetric
reductions from \textbf{4} to \textbf{3} physical bosons can be
performed along $``\phi"$ and/or ``u" directions. They will be
respectively called ``angular" and/or ``radial" reductions.

Let us also redefine the fermionic variables as
\begin{equation}
\psi ^{1A}\equiv \psi ^{A},~\quad \psi ^{2A}\equiv \overline{\psi }^{A}~.
\label{ff}
\end{equation}
Finally, we fix the constant symmetric matrix $A_{ab}$ in (\ref{int}) and (%
\ref{int1}) as follows:
\begin{equation}
A_{12}=A_{21}=M,\quad A_{11}=A_{22}=0.
\end{equation}

In the new variables the ``root'' action given by the sum of (\ref{root3})
and (\ref{int1}) reads\footnote{%
We use the notation $\psi \dot{\bar{\psi}}=\psi ^{A}\dot{\bar{\psi}}%
_{A}$ and similar ones throughout.}
\begin{eqnarray}
S &=&\int dt\left\{ \frac{1}{2}e^{u}G\left( \dot{u}^{2}+\dot{\phi}%
^{2}\right) +2Ge^{u}\frac{\dot{\Lambda}\dot{\bar{\Lambda}}}{1+\Lambda \bar{%
\Lambda}}-\frac{1}{2}Ge^{u}\left( \frac{\Lambda \dot{\bar{\Lambda}}+\dot{%
\Lambda}\bar{\Lambda}}{1+\Lambda \bar{\Lambda}}\right) ^{2}+\right.
\nonumber \\
&&ie^{u}G\frac{\dot{\Lambda}\bar{\Lambda}-\Lambda \dot{\bar{\Lambda}}}{%
1+\Lambda \bar{\Lambda}}\dot{\phi}+\frac{i}{2}G\left( \psi \dot{\bar{\psi}}-%
\dot{\psi}\bar{\psi}\right) -  \nonumber \\
&&\frac{i}{2}\left[ \frac{\dot{\Lambda}\bar{\Lambda}-\Lambda \dot{\bar{%
\Lambda}}}{1+\Lambda \bar{\Lambda}}G_{u}-i\dot{\phi}G_{u}+\dot{\Lambda}%
G_{\Lambda }-\dot{\bar{\Lambda}}G_{\overline{\Lambda }}+i\left( \dot{u}-%
\frac{\dot{\Lambda}\bar{\Lambda}+\Lambda \dot{\bar{\Lambda}}}{1+\Lambda \bar{%
\Lambda}}\right) G_{\phi }\right] \psi \bar{\psi}+  \nonumber \\
&&\frac{i}{2}\left\{ -\frac{\dot{\bar{\Lambda}}}{\left( 1+\Lambda \bar{%
\Lambda}\right) }G_{u}+\left[ \frac{1}{2}\left( 1+\Lambda \bar{\Lambda}%
\right) \left( \dot{u}+i\dot{\phi}\right) +\frac{1}{2}\left( \Lambda \dot{%
\bar{\Lambda}}-\dot{\Lambda}\bar{\Lambda}\right) \right] G_{\Lambda }-\right.
\nonumber \\
&&\left. \frac{i}{2}\left[ \bar{\Lambda}\left( \dot{u}+i\dot{\phi}-\frac{%
\dot{\Lambda}\bar{\Lambda}+\Lambda \dot{\bar{\Lambda}}}{1+\Lambda \bar{%
\Lambda}}\right) +2\dot{\bar{\Lambda}}\right] G_{\phi }\right\} e^{i\phi
}\psi ^{2}-  \nonumber \\
&&\frac{i}{2}\left\{ \frac{\dot{\Lambda}}{\left( 1+\Lambda \bar{\Lambda}%
\right) }G_{u}+\left[ -\frac{1}{2}\left( 1+\Lambda \bar{\Lambda}\right)
\left( \dot{u}-i\dot{\phi}\right) +\frac{1}{2}\left( \Lambda \dot{\bar{%
\Lambda}}-\dot{\Lambda}\bar{\Lambda}\right) \right] G_{\bar{\Lambda}}-\right.
\nonumber \\
&&\left. \frac{i}{2}\left[ \Lambda \left( \dot{u}-i\dot{\phi}-\frac{\dot{%
\Lambda}\bar{\Lambda}+\Lambda \dot{\bar{\Lambda}}}{1+\Lambda \bar{\Lambda}}%
\right) +2\dot{\Lambda}\right] G_{\phi }\right\} e^{-i\phi }\bar{\psi}^{2}-
\nonumber \\
&&\frac{1}{8}e^{-u}\left[ G_{uu}+G_{u}+\left( 1+\Lambda
\bar{\Lambda}\right) ^{2}G_{\Lambda \bar{\Lambda}}+\left(
1+\Lambda \bar{\Lambda}\right) G_{\phi
\phi }+\right.  \nonumber \\
&&\left. \left. i\left( 1+\Lambda \bar{\Lambda}\right) \left( \Lambda
G_{\Lambda \phi }-\bar{\Lambda}G_{\bar{\Lambda}\phi }\right) \right] \psi
^{2}\bar{\psi}^{2}+2Me^{u}\left( -\dot{\phi}+i\frac{\Lambda \dot{\bar{\Lambda%
}}-\dot{\Lambda}\bar{\Lambda}}{1+\Lambda \bar{\Lambda}}+e^{-u}\psi \bar{\psi}%
\right) \right\} ~.  \label{rootact1}
\end{eqnarray}
Accordingly, the transformation properties (\ref{root4}) are
\begin{eqnarray}
\delta \Lambda &=&-\sqrt{1+\Lambda \overline{\Lambda }}e^{-\frac{1}{2}\left(
u-i\phi \right) }\left( \varepsilon ^{A}+\Lambda \bar{\varepsilon}%
^{A}\right) \psi _{A}~,  \nonumber \\
\delta u &=&\frac{1}{\sqrt{1+\Lambda \overline{\Lambda }}}\left[ \left( \bar{%
\varepsilon}^{A}-\bar{\Lambda}\varepsilon ^{A}\right) \psi _{A}e^{-\frac{1}{2%
}\left( u-i\phi \right) }-\left( \varepsilon ^{A}+\Lambda \bar{\varepsilon}%
^{A}\right) \bar{\psi}_{A}e^{-\frac{1}{2}\left( u+i\phi \right) }\right] ~,
\nonumber \\
\delta \phi &=&i\sqrt{1+\Lambda \overline{\Lambda }}\left[ \bar{\varepsilon}%
^{A}\psi _{A}e^{-\frac{1}{2}\left( u-i\phi \right) }+\varepsilon ^{A}\bar{%
\psi}_{A}e^{-\frac{1}{2}\left( u+i\phi \right) }\right] ~,  \nonumber \\
\delta \psi ^{A} &=&-\frac{ie^{\frac{1}{2}\left( u-i\phi \right) }}{\sqrt{%
1+\Lambda \overline{\Lambda }}}\left[ \left( \varepsilon ^{A}+\Lambda \bar{%
\varepsilon}^{A}\right) \left( \dot{u}-i\dot{\phi}-\frac{\Lambda \dot{\bar{%
\Lambda}}+\dot{\Lambda}\bar{\Lambda}}{1+\Lambda \overline{\Lambda
}}\right) +2~\bar{\varepsilon} ^{A}\dot{\Lambda}\right] ,
\label{roottransf}
\end{eqnarray}
with
\begin{equation}
\varepsilon ^{1A}\equiv \varepsilon ^{A}~,\quad \varepsilon ^{2A}\equiv
\overline{\varepsilon }^{A}~.  \label{ff1}
\end{equation}

\subsection{``Angular'' reduction}

In the $\mathcal{N}=4$ superfield formalism this type of reduction is
connected with the existence of composite superfields $V^{ij}$ defined as
\cite{{ikl1},{il}}
\begin{equation}
V^{ij}\equiv \mathcal{Q}{}_{1}^{(i}\mathcal{Q}{}_{2}^{j)}.  \label{lin}
\end{equation}
One can easily check that the ``root'' supermultiplet constraints (\ref
{root1}) imply that the composite superfields $V^{ij}$ (\ref{lin}) obey the
constraints
\begin{equation}
D_{A}^{(i}V^{jk)}=0~.  \label{lin1}
\end{equation}
Being considered as independent $\mathcal{N}=4$ superfields, the constraints
(\ref{lin1}) leave in the superfields $V^{ij}$ the following components:
\begin{equation}
V^{ij},\quad D_{j}^{A}V^{ij},\quad D^{iA}D_{A}^{j}V_{ij}.  \label{lin2}
\end{equation}
Thus, the resulting off-shell component structure is $\mathbf{(3,4,1)}$.
This supermultiplet was introduced in the components in \cite{CR}; the
$\mathcal{N}=4,d=1$ superspace formulation has been presented in
\cite{{is},{BP}}.

By using the parametrization (\ref{newvar}), we get that the first,
physical components of the superfields $V^{ij}$ are the bosonic fields $%
u,\Lambda ,\overline{\Lambda }$, while the angular variable $\phi
$ appears among the components (\ref{lin2}) of the superfields
$V^{ij}$ only through $\dot{\phi}$.

The physical bosonic fields $u,\Lambda ,\overline{\Lambda }$, together with $%
\dot{\phi}$ and the fermionic fields
\begin{equation}
\xi ^{A}\equiv e^{\frac{i}{2}\phi }\psi ^{A}~,\quad \bar{\xi}^{A}\equiv e^{-%
\frac{i}{2}\phi }\bar{\psi}^{A},~  \label{newferm}
\end{equation}
determine the field content of the $\mathcal{N}=4,d=1$ $\mathbf{(3,4,1)}$
linear supermultiplet. This can be seen by inspecting the supersymmetry
transformations. Indeed, by introducing the ``auxiliary'' bosonic field $%
\mathcal{A}$ defined as
\begin{equation}
\mathcal{A}\equiv \dot{\phi}~,  \label{A}
\end{equation}
the supersymmetry transformations (\ref{roottransf}) read
\begin{eqnarray}
\delta \mathcal{A} &=&\partial _{t}\delta \phi =\partial _{t}\left[ i\sqrt{%
1+\Lambda \bar{\Lambda}}e^{-\frac{1}{2}u}\left( \bar{\varepsilon}^{A}\xi
_{A}+\varepsilon ^{A}\bar{\xi}_{A}\right) \right] ~,  \nonumber  \label{tr2}
\\
\delta u &=&-\frac{1}{\sqrt{1+\Lambda \bar{\Lambda}}}e^{-\frac{1}{2}u}\left[
\left( \varepsilon ^{A}+\Lambda \bar{\varepsilon}^{A}\right) \bar{\xi}%
_{A}-\left( \bar{\varepsilon}^{A}-\bar{\Lambda}\varepsilon ^{A}\right) \xi
_{A}\right] ~,  \nonumber \\
\delta \Lambda &=&-\sqrt{1+\Lambda \bar{\Lambda}}e^{-\frac{1}{2}u}\left(
\varepsilon ^{A}+\Lambda \bar{\varepsilon}^{A}\right) \xi _{A}~,  \nonumber
\\
\delta \xi ^{A} &=&-\frac{1}{2}\sqrt{1+\Lambda \bar{\Lambda}}e^{-\frac{1}{2}%
u}\left( \bar{\varepsilon}^{A}\xi _{A}+\varepsilon ^{A}\bar{\xi}_{A}\right) +
\nonumber \\
&&-i\frac{1}{\sqrt{1+\Lambda \bar{\Lambda}}}e^{-\frac{1}{2}u}\left[ \left(
\varepsilon ^{A}+\Lambda \bar{\varepsilon}^{A}\right) \left( \dot{u}-i%
\mathcal{A}-\frac{\dot{\Lambda}\bar{\Lambda}+\Lambda \dot{\bar{\Lambda}}}{%
1+\Lambda \bar{\Lambda}}\right) +2\bar{\varepsilon}^{A}\dot{\Lambda}\right] .
\label{341-lin-susy}
\end{eqnarray}

Let us now consider the reduction of the ``root'' action (\ref{rootact1}).
By analyzing Eq. (\ref{rootact1}), we conclude that the field $\phi $
appears in the action through $\mathcal{A}=\dot{\phi}$ only if the metric $G$
does not depend on $\phi $. By making such an assumption, one may then
replace $\dot{\phi}$ by $\mathcal{A}$ in the action (\ref{rootact1}),
obtaining the component action for the $(\mathbf{3,4,1})$ supermultiplet.

Another immediate consequence of the transformation properties (\ref{tr2})
is the possibility to add to the action (\ref{rootact1}) a FI term of the
form
\begin{equation}
S_{FI}\equiv m\int dt\mathcal{A}\;,  \label{FI}
\end{equation}
where $m$ is an arbitrary, real parameter with physical dimension $[m]=[$%
length$]^{-1}$. It should be noticed that the previously introduced
additional term (\ref{int})-(\ref{int1}) is actually the ``standard'' FI
term for the $\mathbf{(3,4,1)}$ linear supermultiplet (\ref{lin1}), because
the integrand in Eq. (\ref{int}) is nothing but the auxiliary bosonic field $%
D^{iA}D_{A}^{j}V_{ij}$\ . Thus, we conclude that in the considered framework
there are two different FI terms: one is the ``standard'' one expressed by
Eq. (\ref{int}), while the second is given by Eq. (\ref{FI}). The surprising
simplicity of the expression (\ref{FI}) is purely due to the chosen,
convenient parametrization (\ref{newvar}).

By eliminating the ``auxiliary'' field $\mathcal{A}$ through its equation of
motion, the action (\ref{rootact1}) with the FI term (\ref{FI}) added may be
rewritten as
\begin{eqnarray}
S &=&\int dt\left\{ \frac{1}{2}e^{u}G\dot{u}^{2}+2Ge^{u}\frac{\dot{\Lambda}%
\dot{\bar{\Lambda}}}{\left( 1+\Lambda \bar{\Lambda}\right) ^{2}}+\frac{i}{2}%
G\left( \xi \dot{\bar{\xi}}-\dot{\xi}\bar{\xi}\right) \right. +\frac{i}{2}%
\left[ G\frac{\dot{\Lambda}\bar{\Lambda}-\Lambda \dot{\bar{\Lambda}}}{%
1+\Lambda \bar{\Lambda}}-\left( \dot{\Lambda}G_{\Lambda }-\dot{\bar{\Lambda}}%
G_{\bar{\Lambda}}\right) \right] \xi \bar{\xi}  \nonumber \\
&&-\frac{i}{2}\left[ \frac{\dot{\bar{\Lambda}}}{\left( 1+\Lambda \bar{\Lambda%
}\right) }G_{u}-\frac{1}{2}\left( 1+\Lambda \bar{\Lambda}\right) \dot{u}%
G_{\Lambda }\right] \xi ^{2}-\frac{i}{2}\left[ \frac{\dot{\Lambda}}{\left(
1+\Lambda \bar{\Lambda}\right) }G_{u}-\frac{1}{2}\left( 1+\Lambda \bar{%
\Lambda}\right) \dot{u}G_{\bar{\Lambda}}\right] \bar{\xi}^{2}  \nonumber \\
&&+\frac{1}{16}e^{-u}\left\{ G^{-1}\left[ \left( G_{u}+G\right) ^{2}+\left(
1+\Lambda \bar{\Lambda}\right) ^{2}G_{\Lambda }G_{\bar{\Lambda}}%
\right] \left. -2\left[ G_{uu}+G_{u}+\left( 1+\Lambda \bar{\Lambda}\right)
^{2}G_{\Lambda \bar{\Lambda}}\right] \right\} \xi ^{2}\bar{\xi}^{2}\right.
\nonumber \\
&&-\frac{e^{u}}{2G}\left( me^{-u}-2M\right) ^{2}+im\frac{\Lambda \stackrel{%
\cdot }{\bar{\Lambda}}-\dot{\Lambda}\bar{\Lambda}}{1+\Lambda \bar{\Lambda}}+%
\frac{1}{2}\xi \bar{\xi}\left[ me^{-u}\left(
1+\frac{G_{u}}{G}\right)
+2M\left( 1-\frac{G_{u}}{G}\right) \right]  \nonumber \\
&&\left. +\frac{\left( 1+\Lambda \bar{\Lambda}\right) }{4G}\left( G_{\Lambda
}\xi ^{2}-G_{\bar{\Lambda}}\bar{\xi}{}^{2}\right) \left[ me^{-u}-2M\right]
\right\} .  \label{faction1}
\end{eqnarray}

Let us analyze the action (\ref{faction1}).

With $M=m=0$, the action (\ref{faction1}) turns out to describe
the most general $\sigma $-model type component action for the
$\mathcal{N}=4,d=1$ $\mathbf{(3,4,1)}$ linear supermultiplet.
Thus, we may conclude that the kinematical, off-shell relations
between the ``root'' and linear $\mathbf{(3,4,1)}$ supermultiplets
may be prolonged on-shell: indeed, the general $\sigma $-model
type (component) action for the $\mathbf{(3,4,1)}$ linear
supermultiplet is nothing but the ``angular'' reduction of the
$\sigma $-model (component) action for the ``root'' multiplet
(with $M=m=0$).

In the case of non-vanishing $M$ and/or $m$, new interactions are switched
on. It should be noticed that in (\ref{faction1}) the ``standard'' FI term (%
\ref{int}) generates only (properly supersymmetrized) harmonic
oscillator-type potential terms, whereas the FI term defined by Eq. (\ref{FI}%
) generates an interaction with a Dirac monopole.

In order to gain insight about this point, let us consider the bosonic part
of the action (\ref{faction1}) for the trivial metric $G=1$ and with $M=0$.
In such a limit the action reads
\begin{equation}
S_{bos}=\int \!dt\left\{ \frac{1}{2}e^{u}\dot{u}^{2}+2e^{u}\frac{\dot{\Lambda%
}\dot{\bar{\Lambda}}}{\left( 1+\Lambda \bar{\Lambda}\right) ^{2}}-\frac{%
m^{2}e^{-u}}{2}+im\frac{\Lambda \stackrel{\cdot }{\bar{\Lambda}}-\dot{\Lambda%
}\bar{\Lambda}}{1+\Lambda \bar{\Lambda}}\right\} .  \label{bos1}
\end{equation}
The action (\ref{bos1}) may be recognized to be the bosonic part of the $%
\mathcal{N}=4$ superconformal action constructed in \cite{ikl2}, describing
a particle in the field of a Dirac monopole. What is really impressive is
that this action results from the reduction from the \textit{free action}
for the ``root'' supermultiplet. Consequently, for a generic $G$ and $M=0$,
the action (\ref{faction1}) provides us with the same interaction extended
to arbitrary metrics $G\left( \Lambda ,\bar{\Lambda},u\right) $.

Then, we can consider also $G=1$ and $m=0$. In such a case the bosonic part
of the action (\ref{faction1}) reads
\begin{equation}
{\tilde{S}}_{bos}=\int \!dt\left\{ \frac{1}{2}e^{u}\dot{u}^{2}+2e^{u}\frac{%
\dot{\Lambda}\dot{\bar{\Lambda}}}{\left( 1+\Lambda \bar{\Lambda}\right) ^{2}}%
-2M^{2}e^{u}\right\} .  \label{bos2}
\end{equation}
After introducing the new variable
\begin{equation}
x\equiv e^{\frac{u}{2}},
\end{equation}
the action (\ref{bos2}) turns out to describe the bosonic sector of the $%
\mathcal{N}=4$ supersymmetric harmonic oscillator. Therefore, for a generic $%
G$ and $m=0$, the full action (\ref{faction1}) corresponds to the $\mathcal{N%
}=4$ supersymmetrization of the harmonic oscillator-type interaction for
arbitrary metrics $G\left( \Lambda ,\bar{\Lambda},u\right) $.

Thus, we see that the proposed ``angular'' reduction from the ``root''
supermultiplet component action equipped by suitable FI terms, gives rise to
a rather general component action for the $\mathbf{(3,4,1)}$ linear
supermultiplet, with interesting self-interacting terms. It is worth
noticing that the FI term (\ref{FI}), which is trivially simple in the
chosen component parametrization (\ref{newvar}), has instead a rather
complicate description in terms of superfields \cite{ikl2}. By reminding
that there are no general approaches to the construction of potential terms,
the above proposed procedure, which allows one to generate such terms by a
reduction from a ``root'' action, seems to be interesting and useful.

In the next Subsection we move to consider a ``radial'' reduction,
demonstrating that non-trivial potential terms appear in the resulting
reduced action, too.

\subsection{``Radial'' reduction}

The same procedure presented in the previous Subsection can be applied to
eliminate the field $u$ instead of the field $\phi $. This kind of reduction
is related with the existence of a nonlinear $\mathbf{(3,4,1)}$
supermultiplet \cite{{ikl1},{il}}. By introducing new fermionic fields as
\begin{equation}
\eta ^{A}\equiv e^{\frac{u}{2}}\psi ^{A}~,\quad \bar{\eta}^{A}\equiv e^{%
\frac{u}{2}}\bar{\psi}^{A},  \label{rr}
\end{equation}
it turns out that they form, together with the physical bosonic fields $\phi
,\Lambda ,\overline{\Lambda }$ and a new ``auxiliary'' bosonic field $%
\mathcal{B}$ defined as
\begin{equation}
\mathcal{B}\equiv \dot{u},  \label{B}
\end{equation}
a $\mathcal{N}=4,d=1$ supermultiplet. As before, this may be checked by
looking at the supersymmetry transformation properties of such fields.
Indeed, by substituting (\ref{rr}) in (\ref{roottransf}), the following
transformation laws are obtained:
\begin{eqnarray}
\delta \mathcal{B} &=&\frac{d}{dt}\left[ \frac{1}{\sqrt{1+\Lambda \overline{%
\Lambda }}}\left[ \left( \bar{\varepsilon}^{A}-\bar{\Lambda}\varepsilon
^{A}\right) \eta _{A}e^{\frac{i}{2}\phi }-\left( \varepsilon ^{A}+\Lambda
\bar{\varepsilon}^{A}\right) \bar{\eta}_{A}e^{-\frac{i}{2}\phi }\right] %
\right] ~,  \nonumber \\
\delta \phi &=&i\sqrt{1+\Lambda \overline{\Lambda }}\left[ \bar{\varepsilon}%
^{A}\eta _{A}e^{\frac{1}{2}\phi }-\varepsilon ^{A}\bar{\eta}_{A}e^{-\frac{i}{%
2}\phi }\right] ~,  \nonumber \\
\delta \Lambda &=&-\sqrt{1+\Lambda \overline{\Lambda }}e^{\frac{i}{2}\phi
}\left( \varepsilon ^{A}+\Lambda \bar{\varepsilon}^{A}\right) \eta _{A}~,
\nonumber \\
\delta \eta ^{A} &=&-\frac{1}{2}\frac{1}{\sqrt{1+\Lambda \overline{\Lambda }}%
}\left[ \left( \bar{\varepsilon}^{B}-\bar{\Lambda}\varepsilon ^{B}\right)
\eta _{B}e^{\frac{i}{2}\phi }-\left( \varepsilon ^{B}+\Lambda \bar{%
\varepsilon}^{B}\right) \bar{\eta}_{B}e^{-\frac{i}{2}\phi }\right] \eta ^{A}+
\nonumber \\
&&-\frac{ie^{-\frac{i}{2}\phi }}{\sqrt{1+\Lambda \overline{\Lambda }}}\left[
\left( \varepsilon ^{A}+\Lambda \bar{\varepsilon}^{A}\right) \left( \mathcal{%
B}-i\dot{\phi}+\frac{\Lambda \dot{\bar{\Lambda}}+\dot{\Lambda}\bar{\Lambda}}{%
1+\Lambda \overline{\Lambda }}\right) +2\varepsilon ^{A}\dot{\Lambda}\right]
.  \label{tr22}
\end{eqnarray}

For what concerns the corresponding ``radial'' reduction of the action (\ref
{rootact1}), one may notice that the dependence on $u$ may be avoided if a
new metric
\begin{equation}
H\equiv e^{u}G~  \label{HH}
\end{equation}
is introduced, and it is assumed not to depend on $u$, too. Unfortunately,
by looking at the term proportional to $M$ in the last line of the action (%
\ref{rootact1}), one immediately realizes that the interaction term (\ref
{int})-(\ref{int1}) depends on $u$; therefore, in the considered reduction
procedure we necessarily have to disregard it, by putting $A_{(ab)}=0$.
Consequently, by using the definition (\ref{HH}) and assuming the
independence of $H$ on $u$, we may replace $\dot{u}$ with $\mathcal{B}$ in
the action (\ref{rootact1}) with $M=0$, obtaining the $\sigma $-model
component action for the nonlinear $\mathbf{(3,4,1)}$ supermultiplet \cite
{ikl1}.

As for the ``angular'' reduction, one may also add a FI term of the form
\begin{equation}
S_{FI}\equiv \kappa \int dt\mathcal{B}~,  \label{FI2}
\end{equation}
where $\kappa $ is an arbitrary, real parameter with physical dimension $%
[\kappa ]=[$length$]^{-1}$.

Being rewritten in the new variables and after the elimination of
the auxiliary field $\mathcal{B}$, the action (\ref{rootact1}) with
$M=0$ and the FI term (\ref{FI2}) added reads
\begin{eqnarray}
S &=&\int dt\left\{ \frac{1}{2}H\left( \dot{\phi}+i\frac{\dot{\Lambda}\bar{%
\Lambda}-\Lambda \dot{\bar{\Lambda}}}{1+\Lambda \bar{\Lambda}}\right) ^{2}+2H%
\frac{\dot{\Lambda}\dot{\bar{\Lambda}}}{\left( 1+\Lambda \bar{\Lambda}%
\right) ^{2}}-\frac{\kappa ^{2}}{2H}+\frac{i}{2}H\left( \eta \dot{\bar{\eta}}%
-\dot{\eta}\bar{\eta}\right) +\right.  \nonumber \\
&&+\left\{ \frac{1}{2}H\eta \bar{\eta}-\frac{1}{4}\left[
H_{\Lambda }\left( 1+\Lambda \bar{\Lambda}\right) -iH_{\phi
}\bar{\Lambda}\right] e^{i\phi
}\eta ^{2}+\frac{1}{4}\left[ H_{\bar{\Lambda}}\left( 1+\Lambda \bar{\Lambda}%
\right) +iH_{\phi }\Lambda \right] e^{-i\phi }\bar{\eta}^{2}\right\} \dot{%
\phi}+  \nonumber \\
&&+\frac{1}{2}\left[ -\kappa \frac{H_{\phi }}{H}+iH\frac{\dot{\Lambda}\bar{%
\Lambda}-\Lambda \dot{\bar{\Lambda}}}{1+\Lambda \bar{\Lambda}}-i\left( \dot{%
\Lambda}H_{\Lambda }-\dot{\bar{\Lambda}}H_{\bar{\Lambda}}\right) -H_{\phi }%
\frac{\Lambda \dot{\bar{\Lambda}}+\dot{\Lambda}\bar{\Lambda}}{1+\Lambda \bar{%
\Lambda}}\right] \eta \bar{\eta}+  \nonumber \\
&&-\frac{1}{2}e^{i\phi }\left\{ \frac{i}{2}\kappa \left[ \left( 1+\Lambda
\bar{\Lambda}\right) \frac{H_{\Lambda }}{H}-i\bar{\Lambda}\frac{H_{\phi }}{H}%
\right] -iH\frac{\dot{\bar{\Lambda}}}{1+\Lambda \bar{\Lambda}}+\frac{i}{2}%
H_{\Lambda }\left( \dot{\Lambda}\bar{\Lambda}-\Lambda \stackrel{\cdot }{\bar{%
\Lambda}}\right) +\right.  \nonumber \\
&&\left. +H_{\phi }\left( \frac{1}{2}\bar{\Lambda}\frac{\Lambda \stackrel{%
\cdot }{\bar{\Lambda}}+\dot{\Lambda}\bar{\Lambda}}{1+\Lambda \bar{\Lambda}}-%
\dot{\bar{\Lambda}}\right) \right\} \eta ^{2}+  \nonumber \\
&&+\frac{1}{2}e^{-i\phi }\left\{ -\frac{i}{2}\kappa \left[ \left(
1+\Lambda
\bar{\Lambda}\right) \frac{H_{\bar{\Lambda}}}{H}+i\Lambda \frac{H_{\phi }}{H}%
\right] +iH\frac{\dot{\Lambda}}{1+\Lambda \bar{\Lambda}}+\frac{i}{2}H_{\bar{%
\Lambda}}\left( \dot{\Lambda}\bar{\Lambda}-\Lambda \dot{\bar{\Lambda}}%
\right) +\right.  \nonumber \\
&&\left. +H_{\phi }\left( \frac{1}{2}\Lambda \frac{\Lambda \dot{\bar{\Lambda}%
}+\dot{\Lambda}\bar{\Lambda}}{1+\Lambda \bar{\Lambda}}-\dot{\Lambda}\right)
\right\} \bar{\eta}^{2}+  \nonumber \\
&&+\frac{1}{16}\left\{ \frac{1}{H}\left( 1+\Lambda \bar{\Lambda}\right) %
\left[ H_{\phi }^{2}+\left( 1+\Lambda \bar{\Lambda}\right) H_{\Lambda }H_{%
\bar{\Lambda}}+iH_{\phi }\left( \Lambda H_{\Lambda }-\bar{\Lambda}H_{\bar{%
\Lambda}}\right) \right] +\right.  \nonumber \\
&&\left. \left. -2\left( 1+\Lambda \bar{\Lambda}\right) \left[ \left(
1+\Lambda \bar{\Lambda}\right) H_{\Lambda \bar{\Lambda}}+H_{\phi \phi
}+i\left( \Lambda H_{\Lambda \phi }-\bar{\Lambda}H_{\bar{\Lambda}\phi
}\right) \right] \right\} \eta ^{2}\bar{\eta}^{2}\frac{{}}{{}}\right\} .
\label{nla}
\end{eqnarray}

The action (\ref{nla}) provides us with the most general $\sigma $-model
type component action for the nonlinear $\mathbf{(3,4,1)}$ supermultiplet,
with additional specific potential terms. Indeed, one may easily check that
for $H=1$ and $\kappa =0$ the action (\ref{nla}) matches the known $\sigma $%
-model type component action for the $\mathcal{N}=4,d=1$ nonlinear tensor
supermultiplet \cite{ikl1}. As before, we may therefore conclude that the
general $\sigma $-model type action for the $\mathbf{(3,4,1)}$ nonlinear
supermultiplet is nothing but the ``radial'' reduction of the $\sigma $%
-model action for the ``root'' multiplet (with $\kappa =0$).

For non-vanishing $\kappa $, the treatment of the resulting interactions in
the reduced system obtained by ``radial'' reduction is much less clear with
respect to the case of ``angular'' reduction. Indeed, while for the
previously obtained $\mathcal{N}=4,d=1$ linear $\mathbf{(3,4,1)}$
supermultiplet the superfield description of the FI term (\ref{FI}) - although
being complicate - is known \cite{ikl2}, in the case of the $\mathcal{N}%
=4,d=1$ nonlinear $\mathbf{(3,4,1)}$ tensor supermultiplet the
superfield description of the FI term (\ref{FI2}) - which is
simply and automatically introduced by performing the ``radial''
reduction - is instead still unknown.

As previously mentioned, for the linear $\mathbf{(3,4,1)}$
supermultiplet the auxiliary bosonic field actually transforms
only through a full time derivative, and therefore it may be used
for the construction of the corresponding ``standard'' FI term,
which turns out to be nothing but the previously considered
self-interaction terms (\ref{int})-(\ref{int1}). Instead for the
nonlinear $\mathbf{(3,4,1)}$ supermultiplet this fails to happen.

Consequently, an intriguing, unanswered question naturally arising
is the following one: how to introduce and describe interaction
terms for the $\mathcal{N}=4,d=1$ nonlinear $(\mathbf{3,4,1})$
supermultiplet in the superfield formalism?

\setcounter{equation}0

\section{From the ``root'' supermultiplet to the (2,4,2) ones}

As for the previous treatment of the case with \textbf{3} physical bosons,
two $\mathcal{N}=4,d=1$ supermultiplets with \textbf{2} physical bosons are
known; they are the linear and nonlinear chiral supermultiplets (see e.g.
\cite{ikl1}, where the nonlinear one was discovered). While the most general
action for the linear chiral supermultiplet has been known for a long time
\cite{BP}, the general action for its nonlinear chiral counterpart has been
constructed quite recently \cite{NL}.

In this Section we are going to show that the reduction procedure
from the ``root'' action (\ref{rootact1}) works in these cases,
too. Clearly, we do not expect to obtain the most general
potential terms in the reduced systems with \textbf{2} physical
bosons, because in these cases the potential terms may contain
arbitrary functions. Nevertheless, it is rather interesting to
understand which type of potential terms are generated by FI terms
through the reduction procedure.

Due to close analogy with the previous reduction cases, for brevity's sake
our treatment will be rather concise.

\subsection{Nonlinear chiral supermultiplet}

The reduction to this supermultiplet combines both reductions from
the previous Section. Thus, we expect that, after some
redefinition of the components, the physical bosonic fields
$\Lambda ,\overline{\Lambda }$ will form the supermultiplet with
some fermions and the ``auxiliary'' bosonic fields $\dot{u}$ and
$\dot{\phi}$. The suitable fermionic fields may be introduced as
follows:
\begin{equation}
\zeta ^{A}\equiv e^{-\frac{1}{2}\left( u-i\phi \right) }\psi ^{A}~,\quad \bar{%
\zeta}^{A}\equiv e^{-\frac{1}{2}\left( u+i\phi \right)
}\bar{\psi}^{A}.
\end{equation}
By such redefinitions, Eq. (\ref{roottransf}) yields that the supersymmetry
transformations of $\Lambda ,\overline{\Lambda }$, $\mathcal{A}$ (defined in
\ref{A}) and $\mathcal{B}$ (defined in \ref{B}) form a closed set.

In order to reduce the action (\ref{rootact1}), we have to redefine the
metric $G$ as
\begin{equation}
h\equiv e^{u}G\;,
\end{equation}
and assume that $h$ does not depend on $u$ and $\phi $. Finally, we add to
the action (\ref{rootact1}) two FI terms defined as
\begin{equation}
S_{FI}\equiv \int dt\left[ m\mathcal{A}+\kappa \mathcal{B}\right] .
\end{equation}
By eliminating the auxiliary fields $\mathcal{A}$ and $\mathcal{B}$ by their
equations of motion, we get the following action:
\begin{eqnarray}
S &=&\int dt\left\{ 2h\frac{\dot{\Lambda}\dot{\bar{\Lambda}}}{\left(
1+\Lambda \bar{\Lambda}\right) ^{2}}-im\frac{\dot{\Lambda}\bar{\Lambda}%
-\Lambda \dot{\bar{\Lambda}}}{1+\Lambda \bar{\Lambda}}+\frac{i}{2}h\left(
\zeta \dot{\bar{\zeta}}-\dot{\zeta}\bar{\zeta}\right) -\frac{1}{2h}\left(
m^{2}+\kappa ^{2}\right) \right. +  \nonumber \\
&&+\left[ \frac{i}{2}\frac{\dot{\bar{\Lambda}}}{1+\Lambda \bar{\Lambda}}h-%
\frac{i}{4}\left( \kappa +im\right) \left( 1+\Lambda \bar{\Lambda}\right)
\frac{h_{\Lambda }}{h}\right] \zeta ^{2}+  \nonumber \\
&&+\left[ \frac{i}{2}\frac{\dot{\Lambda}}{1+\Lambda \bar{\Lambda}}h-\frac{i}{%
4}\left( \kappa -im\right) \left( 1+\Lambda \bar{\Lambda}\right) \frac{h_{%
\bar{\Lambda}}}{h}\right] \bar{\zeta}^{2}+  \nonumber \\
&&\left. +\frac{i}{2}\left[ \frac{\dot{\Lambda}\bar{\Lambda}-\Lambda \dot{%
\bar{\Lambda}}}{1+\Lambda \bar{\Lambda}}h-\left( \dot{\Lambda}h_{\Lambda }-%
\dot{\bar{\Lambda}}h_{\bar{\Lambda}}\right) \right] \zeta \bar{\zeta}-\frac{1%
}{8}\left( 1+\Lambda \bar{\Lambda}\right) ^{2}\left( h_{\Lambda \bar{\Lambda}%
}-\frac{h_{\Lambda }h_{\bar{\Lambda}}}{h}\right) \zeta ^{2}\bar{\zeta}%
^{2}\right\} .  \label{nlcact}
\end{eqnarray}
Thus, we see that the FI term $\kappa \mathcal{B}$ is responsible only for
potential terms in the action{\ (\ref{nlcact})}, while the term $m\mathcal{A}
$\ generates also an interaction with the magnetic field.

By redefining the metric of the physical bosonic manifold and the fermionic
fields respectively as
\begin{equation}
g\equiv 2\left( 1+\Lambda \overline{\Lambda }\right) ^{-2}h~,~~
\zeta ^{A} \equiv -\sqrt{1+\Lambda \overline{\Lambda }}\psi ^{A}~,
\end{equation}
one may check that the action (\ref{nlcact}) coincides with the
one discussed in \cite{NL}, with the simple holomorphic
superpotential $F\left( \Lambda \right) =(m-i\kappa )\Lambda $.

\subsection{Linear chiral multiplet}

This is the simplest case of reduction, which fully agrees with the
considerations made in \cite{GR}. Let us rename the bosonic variables as
\begin{equation}
q^{11}\equiv z~,\quad q^{22}\equiv \bar{z}~,\quad q^{12}\equiv w~,\quad
q^{21}\equiv -\bar{w}\;.  \label{newqs}
\end{equation}
Consequently, the supersymmetry transformations (\ref{root4}) now read
\begin{equation}
\delta z=-\varepsilon \psi ~,\quad \delta w=-\varepsilon \bar{\psi}~,\quad
\delta \psi ^{A}=-2i\left( \bar{\varepsilon}^{A}\dot{z}+\varepsilon ^{A}\dot{%
\bar{w}}\right) ~.
\end{equation}
Thence, by considering the physical bosonic fields $z$ and $\bar{z}$, the
``auxiliary'' bosonic fields
\begin{equation}
\mathcal{C}\equiv \dot{w},~~\mathcal{\bar{C}}\equiv \dot{\bar{w}},
\label{mbox}
\end{equation}
and the fermionic fields $\psi ^{A}$ and $\bar{\psi}{}^{A}$ defined in (\ref
{ff}), it is clear that all together they form a close multiplet.

By substituting the definitions (\ref{newqs}) and (\ref{mbox}) in the action
(\ref{root3}) and assuming the independence of the metric $G$ on $w$ and $%
\bar{w}$, one obtains the $\sigma $-model component action for the
$\left( \mathbf{2,4,2}\right) $ linear chiral multiplet:
\begin{eqnarray}
S &=&\int dt\left\{ 2G\left( \dot{z}\dot{\bar{z}}+\mathcal{C\bar{C}}\right) +%
\frac{i}{2}G\left( \psi \dot{\bar{\psi}}-\dot{\psi}\bar{\psi}\right) -\frac{i%
}{2}\left[ \left( G_{z}\dot{z}+G_{\bar{w}}\mathcal{\bar{C}}-G_{\bar{z}}\dot{%
\bar{z}}-G_{w}\mathcal{C}\right) \psi \bar{\psi}+\right. \right.  \nonumber
\\
&&+\left. \left. \left( G_{z}\mathcal{C}-G_{\bar{w}}\dot{\bar{z}}\right)
\psi ^{2}-\left( G_{\bar{z}}\mathcal{\bar{C}}-G_{w}\dot{z}\right) \bar{\psi}%
^{2}\right] -\frac{1}{8}\left( G_{z\bar{z}}+G_{w\bar{w}}\right) \psi ^{2}%
\bar{\psi}^{2}\right\} .
\end{eqnarray}

Thus, after adding to the action (\ref{root3}) two FI terms defined as
\begin{equation}
S_{FI}\equiv \int dt\left[ ~\mu~ \mathcal{C}+\bar{\mu}~\mathcal{\bar{C}~}%
\right]  \label{FI-FI}
\end{equation}
(where $\mu $ is an arbitrary, complex parameter with physical dimension $%
[\mu ]=[$length$]^{-1}$), we may eliminate the ``auxiliary'' bosonic fields $%
\mathcal{C}$ and $\mathcal{\bar{C}}$ by using their equations of motion. The
final result reads
\begin{eqnarray}
S &=&\int dt\left\{ 2G\dot{z}\dot{\bar{z}}+\frac{i}{2}G\left( \psi \dot{\bar{%
\psi}}-\dot{\psi}\bar{\psi}\right) -\frac{\mu \bar{\mu}}{2G}-\frac{i}{2}%
\left( G_{z}\dot{z}-G_{\bar{z}}\dot{\bar{z}}\right) \psi \bar{\psi}+\right.
\nonumber \\
&&+\left. \frac{i}{4}\left( \bar{\mu}~\frac{G_{z}}{G}\psi ^{2}-\mu ~\frac{G_{%
\bar{z}}}{G}\bar{\psi}^{2}\right) -\frac{1}{8}\left( G_{z\bar{z}}-\frac{%
G_{z}G_{\bar{z}}}{G}\right) \psi ^{2}\bar{\psi}^{2}\right\} ~,
\end{eqnarray}
and corresponds to the on-shell, general $\sigma $-model type component
action for the $\left( \mathbf{2,4,2}\right) $ linear chiral multiplet \cite
{GR} with peculiar potential terms \textbf{.\smallskip \medskip }

Finally, it should be mentioned that one may go further, and
perform also the reduction to the $N=4,d=1$\ $(\mathbf{1,4,3})$\
``old'' tensor supermultiplet \cite{ikl1}. This reduction goes in
a similar way, producing the known action \cite{leva}. Therefore,
we will not discuss this case here.\setcounter{equation}0

\section{Conclusion}

Being motivated by the relations between different $\mathcal{N}=4$
supermultiplets in $d=1$ dimensions established by J.~Gates, Jr. and
L.~Rana, in this paper we considered the $\mathcal{N}=4,d=1$ supermultiplet with $(%
\mathbf{4,4,0})$ component content as the \textbf{``}root'' one for all
other known (linear and nonlinear) supermultiplets of $\mathcal{N}=4$
supersymmetric mechanics.

Starting from the most general sigma-model type action for such a multiplet,
we explicitly demonstrated that the actions for the rest of linear, and two
nonlinear, $\mathcal{N}=4,d=1$ supermultiplets can be easily obtained by a
proper reduction procedure.

One of the intrinsic peculiarities of the reduction scheme is the
possibility to introduce Fayet-Iliopoulos terms. Such terms give
rise to potential terms and in some cases to terms describing the
interaction with a magnetic field, too.

We demonstrate that known $\mathcal{N}=4$ superconformal actions,
together with their possible interactions, appear as results of
the reduction from a \textit{free action} for the ``root''
supermultiplet. As a byproduct, we also construct an
$\mathcal{N}=4$ supersymmetric component action for the linear
$(\mathbf{3,4,1})$ supermultiplet, containing both an interaction
with a Dirac monopole and a (properly supersymmetrized) harmonic
oscillator-type potential, generalized for arbitrary conformally
flat metrics.

Thus, the main conclusion is that in $\mathcal{N}=4,d=1$ supersymmetry there
exists a ``root'' supermultiplet with a related ``root'' action from which,
by a proper reduction, one may construct the rest of the supermultiplets and
the corresponding actions.

One of the most serious and intriguing questions arising in such a
framework concerns the possibility to apply the same reduction
procedure to intrinsically nonlinear $\mathcal{N}=4,d=1$
supermultiplets, recently found to be crucial in order to obtain
supersymmetric actions with non-conformally flat bosonic manifolds
\cite{EH}. Up to now, we dealt only with supermultiplets which can
be described in terms of nonlinear realizations of the unique $%
\mathcal{N}=4,d=1$ superconformal group $D\left( 2,1;\alpha \right) $ \cite
{ikl1}. At present, not too much is known about the geometric origin of the
intrinsically nonlinear supermultiplets of $\mathcal{N}=4$ supersymmetric
mechanics. So this problem remains open.

Another promising direction of research concerns the study of $\mathcal{N}>4$%
-extended supersymmetric systems; in this respect, particularly interesting
appears the generalization of the proposed reduction to the case $\mathcal{N}%
=8$. So far, a detailed investigation of the superfield structure
of all linear $N=8, d=1$ supermultiplets was carried out in
\cite{ABC}, whereas new variants of $N=8$ supersymmetric mechanics
were developed in \cite{{bkn1},{bks1},{sut},{bis}}. Going beyond
$\mathcal{N}=4$, one usually gets that the metric of the bosonic
manifold is much more restrictive with respect to the
$\mathcal{N}=4$ case, and therefore one may expect that some
modification of the reduction scheme proposed above will be
needed. Moreover, it is well known that there are four different
$\mathcal{N}=8,d=1$ superconformal groups \cite{VP}. Thus, in
$\mathcal{N}=8$ supersymmetric mechanics it is not completely
clear whether relations between different supermultiplets having a
natural description within different superconformal groups may
exist at all. There exists also an $\mathcal{N}=8,d=1$
supermultiplet with an infinite number of auxiliary components; a
rather ambitious task would be to establish some
relations (if any) between finite and infinite-dimensional $\mathcal{N}%
=8,d=1 $ multiplets.

It is worth noticing once again that the interaction with a Dirac
monopole, obtained in the component action of the
$(\mathbf{3,4,1})$ linear supermultiplet by performing an
``angular'' reduction from the suitably parameterized ``root''
$(\mathbf{4,4,0})$ component action, appears as the result of
adding the simplest Fayet-Iliopoulos term (\ref{FI}) to the
``root'' action (\ref{rootact1}). Consequently, it might happen
that in the case of (possibly existing) $\mathcal{N}=8$ ``root''
supermultiplet(s) a similar mechanism produces an interaction with
a Yang monopole. Of course, it is desirable at least to check such
a possibility.\footnote{Issues related to
the possibility to exactly solve extended supersymmetric mechanics systems
after inclusion of a constant magnetic field
were treated for $N=4$ in earlier papers \cite{BN,BNY}. From an historical perspective,
we wish to recall the early work in \cite{halp}, discussing the $\mathcal{N}=4$
and $\mathcal{N}=16$ cases.}

Finally, it is clear that all the above results, here obtained in the
component formalism, may be reformulated in terms of superfields. It would
be very nice to understand how the proposed reduction procedure works in
superspace, and in particular how the introduction of interactions by
suitable Fayet-Iliopoulos terms may be described in the superfield formalism.

\section*{Acknowledements}

We are gratefully obliged to Jim Gates, Evgeny Ivanov and Armen Nersessian for
valuable discussions. This research was
partially supported by the European Community's Marie Curie
Research Training Network under contract MRTN-CT-2004-005104
Forces Universe, and by the
INTAS-00-00254 grant.

\end{document}